\documentstyle[preprint,tighten,eqsecnum,aps]{revtex}
\begin{document}
\draft
\preprint{UTPT-94-01}
\title{Nonsymmetric Gravity Does Have Acceptable Global Asymptotics.}
\author{ Neil J. Cornish and John W. Moffat}
\address{Department of Physics, University of Toronto \\ Toronto,
Ontario M5S 1A7, Canada}
\maketitle
\begin{abstract}
We consider the claim by Damour, Deser and McCarthy \cite{DDM3}
that nonsymmetric gravity theory has unacceptable global asymptotics. We
explain why this claim is incorrect.
\end{abstract}
\narrowtext

\section{Introduction}
In a series of papers \cite{DDM1,DDM2,DDM3}, Damour, Deser and McCarthy
(DDM) have claimed
that the nonsymmetric gravitational theory (NGT) is theoretically inconsistent.
They noted that a spurious ``gauge invariance'' in the linearised version
of NGT did not generalise to curved spacetime, from which they concluded
that ghost excitations would occur. However, the ``gauge invariance'' in
question is simply an artifact of the linearisation, and plays no role in
ensuring the conservation of the true Noether charges of the theory. The loss
of such an invariance is as unimportant as its existence.

They went on to argue that generic solutions suffered unacceptable asymptotic
behaviour at future null infinity, $\cal{I}^{+}$. This argument was based
on the supposed behaviour of a Lagrange multiplier field which, in fact, can
and should be eliminated from the field equations. We explained the pitfalls
of such a treatment in ref.\cite{CorMoff}, and we shall expand on our
original explanation in this note. It has since been shown that general
radiative solutions can be found with good asymptotic behaviour at
$\cal{I}^{+}$ \cite{pla,CMT}.

Damour, Deser and McCarthy now accept that good asymptotic behaviour can
be found at $\cal{I}^{+}$, but claim this can only be achieved at the
expense of bad asymptotic behaviour at $\cal{I}^{-}$ \cite{DDM3}. They
base this claim on a lemma which states that any solution of an
inhomogeneous wave equation which falls off faster than $1/r$ at
$\cal{I}^{+}$ must be an advanced solution. We point out in this note
that the lemma they use is not applicable to the system of equations being
studied, since they apply Green's theorem to a hyperbolic system
while neglecting lower order differential constraints on the
boundary of integration. Consequently, their assertion that NGT has bad
global asymptotics is unfounded.

\section{Analysis of the Field Equations}
For ease of comparison with the work of DDM, we shall adopt their
unconventional notation for the field variables. Performing an expansion
in powers of the anti-symmetric field $B_{\mu\nu}$ about a fixed symmetric
background $G_{\mu\nu}$, the NGT field equations \cite{Moff79} read to
first order:
\begin{eqnarray}
R_{(\mu\nu)}(G)&=& 0 \; , \\
\nabla^{\alpha}\nabla_{\alpha}B_{\mu\nu}
-2R^{\alpha\;\beta}_{\;\mu\;\nu}B_{\alpha\beta}
&=&{2\over 3}(\partial_{\mu}\Gamma_{\nu}-
\partial_{\nu}\Gamma_{\mu})\; , \label{mix} \\
\nabla^{\nu}B_{\mu\nu}&=&0 \; , \label{sdiv}
\end{eqnarray}
which may be supplemented by the gauge choice:
\begin{equation}
\nabla^{\mu}\Gamma_{\mu}=0 \; .
\end{equation}
The covariant derivative, $\nabla$, is that of the background metric
$G_{\mu\nu}$, and the vector $\Gamma_{\mu}$ is a non-dynamical Lagrange
multiplier. This Lagrange multiplier appears in the linearised NGT Lagrangian
density via the term:
\begin{equation}
{\cal L}_{\Gamma}=\sqrt{-G}B^{\mu\nu}(\partial_{\nu}\Gamma_{\mu}
-\partial_{\mu}\Gamma_{\nu})\; ,
\end{equation}
which gives rise to the field equation (\ref{sdiv}). In the language of field
theory, we see that $\Gamma_{\nu}$ does not have a propagator, since
the Lagrangian cannot be formulated to have a kinetic energy term for
$\Gamma_{\nu}$. As such, it makes no sense to talk
about $\Gamma_{\nu}$ having retarded or advanced propagator solutions.
We shall return to this crucial point later.

In order to solve the field equations, we need to take the divergence and
cyclic curl of (\ref{mix}), while remembering that the solutions to these
higher derivative equations are constrained to be solutions
to the primary lower derivative equation, (\ref{mix}). The equations for
$B_{\mu\nu}$ and $\Gamma_{\mu}$ then read
\begin{eqnarray}
\nabla^{\nu}B_{\mu\nu}&=&0 \; , \label{sd} \\
\nabla^{\alpha}\nabla_{\alpha}B_{ \{ \mu\nu ,\kappa \} }
-2\nabla_{\{\kappa}(R^{\alpha\;\beta}_{\;\mu\;\nu\} }B_{\alpha\beta})
+\nabla_{\alpha}(R^{\alpha\beta}_{\;\;\;\{\kappa\mu}B_{\nu\}\beta})&=&0
\; , \label{cc} \\
\nabla^{\alpha}\nabla_{\alpha}\Gamma_{\nu}&=&
-3\nabla^{\mu}(R^{\alpha\;\beta}_{\;\mu\;\nu }B_{\alpha\beta})
\; . \label{yuk}
\end{eqnarray}
The first two sets of equations, (\ref{sd}, \ref{cc}), represent six equations
for the six $B_{\mu\nu}$. These six equations {\em fully determine
$B_{\mu\nu}$ with no reference to the Lagrange multiplier $\Gamma_{\nu}$},
which is what one expects from a system of equations with a
Lagrange multiplier. The last equation, (\ref{yuk}), can be used to solve
for the Lagrange multiplier $\Gamma_{\nu}$ {\em once the six $B_{\mu\nu}$
are known}. We note that on its own, (\ref{yuk})
can only determine the LHS of (\ref{mix}) to be $2/3(\nabla_{\mu}\Gamma_{\nu}-
\nabla_{\nu}\Gamma_{\mu})+F_{\mu\nu}$ where $F_{\mu\nu}$ is any skew
tensor that satisfies $\nabla^{\nu}F_{\mu\nu}=0$.

In their analysis, DDM concentrate their attention on first solving for
the Lagrange multiplier $\Gamma_{\nu}$ via equation (\ref{yuk}). This
approach is fraught with problems. Firstly, the $\nabla RB$ ``source term''
is an unknown function unless you have already solved for $B_{\mu\nu}$.
DDM fail to check whether their eventual solution for $B_{\mu\nu}$ is a
self-consistent solution of this equation. It is not.
Secondly, the hyperbolic differential operator $\nabla^{\alpha}
\nabla_{\alpha}$ in (\ref{yuk}) demands propagating, retarded and advanced
$1/r$ Green's function solutions for $\Gamma_{\nu}$. This leads to a distorted
physical picture, as the primary field equation, (\ref{mix}), for
$\Gamma_{\nu}$ is not a wave equation. Wave solutions for $\Gamma_{\mu}$ play
no part in the physics of NGT.

Additionally, DDM base their arguments on the Green's function for the
flat-space d'Alembertian $\Box$ instead of the operator
$\nabla^{\alpha} \nabla_{\alpha}$, while at the same time keeping the source
term $\nabla RB$. This treatment is inconsistent since
$(\nabla^{\alpha}\nabla_{\alpha}-\Box)\Gamma \, \sim {\cal O}(R\Gamma)$
is of the same order as the source term.

We shall see that the most important of these errors is their failure to
correctly treat the spurious solutions for $\Gamma_{\nu}$, which are
manufactured in taking the divergence of the field equations.

\section{The explicit faults in DDM's argument}

While the reasons given above more than suffice to invalidate the
proof given by DDM, it is instructive to see the failings of their
arguments shown explicitly. Indeed, one needs to look no further
than the static case to illustrate the errors in their analysis.

For static systems, the uniqueness theorem for inhomogeneous wave
equations cited by DDM \cite{fock} reduces to the statement that
$\Gamma_{\mu}$ will fall-off as $1/r$ at spatial or null infinity.

Explicitly, we find for the component $\Gamma_{t}$ that (\ref{yuk})
becomes
\begin{equation}
\left[\nabla^{2}-{2M \over r^2}\left(r{\partial^{2} \over
\partial r^{2} }+2{\partial \over \partial r}\right)\right]\Gamma_{t}
=3\nabla^{\mu}(R^{\alpha\;\beta}_{\;\mu\; t }B_{\alpha\beta}) \label{G} \; ,
\end{equation}
where $\nabla^{2}$ is the usual flat-space Laplacian for a scalar
field and $M$ is the mass associated with the Schwarzschild background.
If we follow the argument of DDM - by neglecting the fact that the
source term is unknown and the operator acting on $\Gamma_{t}$
is not just the flat-space Laplacian - then we conclude from the
uniqueness theorem that $\Gamma_{t} \sim 1/r$.

If we then do as DDM suggest, and feed this information into (\ref{mix})
we find
\begin{equation}
\left[\left(\nabla^2+{2 \over r}{\partial \over \partial r}
+{2 \over r^2}\right)-{2M \over r}\left({\partial^{2} \over
\partial r^{2} }+{3 \over r}{\partial \over \partial r}+{4 \over r^2}
\right)\right]B_{tr}={2 \over 3}{\partial\Gamma_{t} \over \partial r}
\sim {1/r^2} \; , \label{s}
\end{equation}
from which we conclude $B_{tr} \sim r^{0}$, exactly as predicted by DDM.

The above analysis would seem to provide a perfect example of DDM's
claim that NGT has unacceptable asymptotic behaviour.
Let us now look at what has been forgotten.  Firstly, we have neglected
the field equation (\ref{sd}) which demands
\begin{equation}
\nabla^{\alpha}B_{t\alpha}=0 \; , \label{bang you're dead}
\end{equation}
so that $B_{tr} \sim r^{-2}$. DDM's putative solution, $B_{tr} \sim r^{0}$,
{\em is not a solution to} (\ref{bang you're dead}). Putting
$B_{tr} = l^2/r^{2}$ (where $l^{2}$ is a constant of integration)
into (\ref{s}) gives
\begin{equation}
{-8Ml^2 \over r^5} \sim {1 \over r^2}\; ,
\end{equation}
which  clearly excludes the $1/r$ behaviour for $\Gamma_{t}$ suggested by
(\ref{G}). This is simply an example of a higher order derivative equation,
(\ref{G}), having solutions which are incompatible with the lower order
equation, (\ref{s}), from which it came. Restoring the term
$\partial_{r}\Gamma_{t}$ in (\ref{s}), we
see that the true solution for $\Gamma$ has $\Gamma_{t} =3Ml^2 /r^{4}$.

The above calculation has now faithfully reproduced the exact NGT static
solution to first order in $B_{\mu\nu}$ - a solution which represents a
non-trivial extension of the Schwarzschild solution of General Relativity (GR)
with {\em good asymptotic behaviour for} $B_{\mu\nu}$.
The putative $1/r$ Green's function solution for $\Gamma$ was simply an
artifact produced by taking additional derivatives of the field equations.

Another way of seeing what went wrong with DDM's argument is to recall that
we are free to add a tensor of integration, $F_{\mu\nu}$,
to the LHS of (\ref{s}) , where $F_{\mu\nu}$ is
a solution of $\nabla^{\alpha}F_{t \alpha}=0$.
In this case, we may add $F_{tr}
\sim 1/r^{2}$ to the LHS of (\ref{s}) with the coefficient chosen to eliminate
the $1/r^2$ term given by the curl of $\Gamma_{t}$. In this way,
$B_{tr}$ is no longer driven to behave as $r^{0}$. From this argument we see
that the putative $1/r$ solution for $\Gamma_{t}$ was a purely homogeneous
solution which must be chosen to vanish.

Generalising our analysis to the time-dependent case we find a very similar
picture. Again the higher-derivative equation for $\Gamma$ admits
homogeneous solutions with $1/r$ fall-off, while the lower order constraints
on $\Gamma$ demand that these solutions be discarded.

We shall demonstrate this in the context of wave solutions
on a radiative, axi-symmetric GR background\cite{pla}. To leading order, the
GR background is described by $M(u,\theta)$ and $c(u,\theta)$, where $u=t-r$
is retarded time, the mass associated with the background is given by the
angular average of $M$ and the time rate of change of this mass is given by
the angular average of $-(\partial_{u}c)^2$. We shall only sketch the main
steps in solving the equations, as the full solution is derived in detail
in ref.\cite{pla}.

Beginning with the wave equations for $\Gamma_{\mu}$ we find
\begin{equation}
\Box \Gamma_{u}+{\cal O}(R\Gamma)=-3\nabla^{\mu}(R^{\alpha\;\beta}_{\;\mu\; u }
 B_{\alpha\beta}) \; ,
\end{equation}
where $\Box$ is the usual flat-space d'Alembertian for a scalar field, and the
extra background terms of order $R\Gamma$ are given by
\begin{eqnarray}
{\cal O}(R\Gamma)&=&{2 \over r}\left[M{\partial^{2}\Gamma_{u} \over \partial
r^2}+{\partial M \over \partial u}{\partial \Gamma_{r} \over \partial r}\right]
+{2 \over r^2}\left[\left(4M+2c+2c{\partial c \over \partial u}-
{\partial^2 c \over \partial \theta^{2}}-3\cot\theta{\partial c \over
\partial \theta}\right){\partial \Gamma_{u} \over \partial r} \right. \nonumber
\\
&& +c^2{\partial^2 \Gamma_{u} \over \partial r \partial u}-{c^2 \over 2}
{\partial^{2} \Gamma_{u} \over \partial r^2}
 -2\left({\partial^2 c \over \partial \theta \partial u}
+2\cot\theta{\partial c \over \partial u}\right){\partial \Gamma_{\theta}
\over \partial r} \nonumber \\
&& \left. -2M{\partial \Gamma_{r} \over \partial u}
-2\left({\partial c \over \partial u}\right)^{2}\Gamma_{r}
-c{\partial c \over \partial u}{\partial \Gamma_{r} \over \partial r} \right]
+ \dots \; .
\end{eqnarray}
Although it is inconsistent to drop these terms while keeping the $RB$
source term, we shall follow the method proposed by DDM and drop them anyway.
The wave equation for $\Gamma_{r}$
then reads
\begin{equation}
\left[\Box-{2\over r}{\partial \over \partial r}-{2 \over r^2}
+{2 \over r}{\partial \over \partial t}\right]\Gamma_{r}+\left[{2 \over r}
{\partial \over \partial r}+{2 \over r^2}\right]\Gamma_{u}=
-3\nabla^{\mu}(R^{\alpha\;\beta}_{\;\mu\; r}B_{\alpha\beta}) \; .
\end{equation}
When combined with the gauge condition $\nabla^{\alpha}\Gamma_{\alpha}=0$
and the wave equation for $\Gamma_{\theta}$, we find that the above equations
have the usual retarded wave solutions $\Gamma_{u}=f(u,\theta)/r$,
$\Gamma_{r}=g(u,\theta)/r^2$ and $\Gamma_{\theta}=h(u,\theta)$ (in orthonormal
coordinates this means $\Gamma_{{\hat{\theta}}} \sim 1/r$). Since the source
terms are at present unknown, we cannot give explicit forms for $f$, $g$ and
$h$, although the gauge condition does demand
$f+\partial_{u}g=\partial_{\theta}
h+h\cot\theta$.

Turning to the $(ur)$ component of (\ref{mix}) we find
\begin{eqnarray}
\left[\Box-{2\over r}{\partial \over \partial r}-{2 \over r^2}
-{2 \over r}{\partial \over \partial t}\right]B_{ur}+{\cal O}(RB)
&=&{2 \over 3}\left({\partial \Gamma_{r} \over \partial u}
-{\partial \Gamma_{u} \over \partial r}\right)
\nonumber \\
&=&{2\partial_{\theta}\left(h\sin\theta\right) \over 3r^2\sin\theta }\; .
\label{neat}
\end{eqnarray}
 From this we would conclude $B$ has the unacceptable asymptotic form
$B_{ur} \sim r^{0}$, as promised by DDM.
However, this is in conflict with the field equations (\ref{sd})
and (\ref{cc}) which demand to leading order that $B_{ur}=l^2(u,\theta)/r^2$.
The lower order field equations for $\Gamma_{\nu}$ again demand that the
$1/r$ solution for $\Gamma$ be dropped, as was the case for static
solutions. In this case we require $\partial_{\theta}(h\sin\theta)=0$ which
implies $h=0$ to ensure regularity on the polar axis.
Notice that setting $h(u,\theta)=0$ removes the only transverse, propagating
component of $\Gamma_{\nu}$, and that the remaining components of
$\Gamma_{\nu}$
are longitudinal and non-propagating, despite the $1/r$ fall-off for
$\Gamma_{u}$. Such important subtleties are lost in DDM's analysis, for they
treat $\Gamma_{\nu}$ as a scalar. It is also worth noting from the full
solution\cite{pla} that the one transverse component of $B_{\mu\nu}$,
$B_{u\theta}$, does have a $1/r$ retarded wave solution. The physical fields,
$B_{\mu\nu}$, that we expect to propagate do, and the non-dynamical Lagrange
multiplier fields, $\Gamma_{\nu}$, do not. We would be surprised if we had
found otherwise.

As in the static case, we see that the extra solutions for
$\Gamma_{\nu}$, which are generated by taking additional derivatives of the
field equations, can be cancelled by the tensor of integration $F_{\mu\nu}$.
Another way of seeing this is to recognise that a consistent
treatment of the ``inhomogeneous wave equation'' for $\Gamma$ demands that
we drop the source term $\nabla RB$ when using the flat-space d'Alembertian
for $\Gamma$. Then we see that the solutions for $\Gamma$ which are independent
of the background parameters $M$ and $c$ are simply homogeneous solutions
which we are free to discard.

The above collection of results allows us to address DDM's lemma head-on.
They claim that the fast fall-off of $\Gamma$ at ${\cal I}^{+}$ comes about
because our solutions for $\Gamma$ are advanced solutions. This is clearly not
the case, as our solutions for $\Gamma$ are explicitly retarded solutions
- despite the fact they do not describe transverse waves with $1/r$ fall-off.
If we accept DDM's contention that $\Gamma$ is described by advanced Green's
functions, we find this leads to a contradiction. If $\Gamma$, and hence
$B$, are functions of advanced time, $v=t+r$, their lemma demands
that $\Gamma$ has propagating, $1/r$, solutions at ${\cal I}^{-}$.
By a simple time reversal of our solutions in terms of retarded time, with
the added simplification that the GR background is static at ${\cal I}^{-}$,
we find from the wave equation for $\Gamma$ that
$\Gamma_{{\hat{\theta}}}=h(v,\theta)/ r +{\cal O}(1/r^2)$, while
the lower order equation (\ref{mix}), and the fact that
$B_{vr}=l^2(v,\theta)/r^2$, demand $h(v,\theta)=0$. This proves by
contradiction that DDM's lemma is not valid for $\Gamma$.

In their latest paper \cite{DDM3}, DDM go on to repeat the same flawed analysis
they applied to the Lagrange multiplier $\Gamma$, by taking two derivatives of
(\ref{mix}) in order to discuss a fourth order wave equation for $B_{\mu\nu}$.
Again, the original, lower order field equations do not allow the badly behaved
solutions of the higher order equations.
All of the pitfalls that plague DDM's analysis can be avoided, if one
works from the outset with (\ref{sd}, \ref{cc}) and uniquely solves for
$B_{\mu\nu}$ without reference to the Lagrange multiplier $\Gamma$.
This was the method first employed by Einstein and Straus when solving an
analogous system of equations in Unified Field Theory \cite{Al}. Using the same
method, we have found exact, radiative solutions, well behaved at
${\cal I}^{+}$ and ${\cal I}^{-}$ \cite{pla,CMT}. These solutions represent
a non-trivial modification to the GR limit of this system and lead to
the prediction that the quadrupole moment of a source will decrease more
rapidly in NGT than GR.

To summarise, we have shown that DDM's claim that NGT has bad global
asymptotics is invalid, since it is based on wave solutions for
a Lagrange multiplier field which fail to solve the original, lower
order field equations. This invalidates the use of their lemma, since
it can only be used for fields with lower order constraints,
when those constraints do not demand that the propagating modes vanish
(as is the case in electromagnetism, for example).
If these constraints are properly accounted for, we find that NGT has
non-trivial radiative solutions with good global asymptotics.

\section*{Acknowledgements}
This work was supported by the Natural Sciences and Engineering
Research Council of Canada. One of the authors (NJC) is
grateful for the support provided by a Canadian Commonwealth
Scholarship. We thank M. Clayton and P. Savaria for helpful
discussions.


\begin{references}
\bibitem{DDM1} Damour T., Deser S. \& McCarthy J.,
{\em Phys. Rev.} D {\bf 45}, R3289, 1992.
\bibitem{DDM2} Damour T., Deser S. \& McCarthy J., {\em Phys. Rev.} D
{\bf 47}, 1541, 1993.
\bibitem{DDM3} Damour T., Deser S. \& McCarthy J., Multi-preprint
IHES/P/93/56, BRX TH-353, IASSNS-HEP-93/67, ADP-93-221/M20, gr-qc/9312030,
1993.
\bibitem{CorMoff} Cornish N. J. \& Moffat J. W.,
{\em Phys. Rev.} D {\bf 47}, 4421, 1993.
\bibitem{pla} Cornish N. J., Moffat J. W. \& Tatarski D. C.,
{\em Phys. Lett.} A {\bf 173}, 109, 1993.
\bibitem{CMT} Cornish N. J., Moffat J. W. \& Tatarski D. C.,
University of Toronto Preprint UT-PT-92-17, gr-qc/9306033, 1992.
\bibitem{Moff79} Moffat J. W., {\em Phys. Rev.} D {\bf 19}, 3554,
1979; {\em Phys. Rev.} D {\bf 35}, 3733, 1987.
\bibitem{fock} Fock V., {\it The Theory of Space, Time and Gravitation},
(Pergamon Press, Oxford, 1966); see section 92.
\bibitem{Al} Einstein A. and Straus E. G., {\em Ann. Math.} {\bf 47}, 731
(1946).
\end{references}
\end{document}